\documentclass[epj]{svjour}
\usepackage{amsfonts,amssymb,amsbsy,amsmath}
\usepackage{mathrsfs}
\usepackage{pdfpages}
\usepackage{graphicx,graphics,epsfig,rotating,slashed}
\usepackage{color,pstricks}

\def\lsim{\mathrel{\rlap{
\lower4pt\hbox{\hskip-3pt$\sim$}}
    \raise1pt\hbox{$<$}}}     %less than approx. symbol
\def\gsim{\mathrel{\rlap{
\lower4pt\hbox{\hskip-3pt$\sim$}}
    \raise1pt\hbox{$>$}}}     %greater than or approx. symbol

\begin{document}
\title{Supporting the search for the CEP location with nonlocal PNJL models constrained by Lattice QCD}
\author{Gustavo~A.~Contrera\inst{1,2,3}
%\email{contrera@fisica.unlp.edu.ar}
%\email{ag.grunfeld@gmail.com}
\thanks{e-mail: contrera@fisica.unlp.edu.ar},
A.~Gabriela~Grunfeld\inst{3,4}  \and David ~Blaschke\inst{5,6,7}
}                     % Do not remove
\institute{IFLP, UNLP, CONICET, Facultad de Ciencias Exactas, calle 49 y 115, La Plata, Argentina
\and
Gravitation, Astrophysics and Cosmology Group, FCAyG, UNLP, La Plata, Argentina
\and
CONICET, Rivadavia 1917, 1033 Buenos Aires, Argentina
\and
Departamento de F\'\i sica, Comisi\'on Nacional de Energ\'{\i}a At\'omica, (1429) Buenos Aires, Argentina
\and
Institute for Theoretical Physics, University of Wroclaw, 50-204 Wroclaw, Poland
\and
Joint Institute for Nuclear Research, 141980 Dubna, Moscow Region, Russia
\and
National Research Nuclear University (MEPhI), 115409 Moscow, Russia}

\date{Received: date / Revised version: date}
\abstract{
We investigate the possible location of the critical endpoint in the QCD phase
diagram based on nonlocal covariant PNJL models including a vector interaction channel.
The form factors of the covariant interaction are constrained by lattice QCD data for the quark propagator.
The comparison of our results for the pressure including the pion contribution and the scaled pressure shift
$\Delta P / T^4$ vs $T/T_c$ with lattice QCD results shows a better agreement when Lorentzian formfactors for the nonlocal interactions and the wave function renormalization are considered.
The strength of the vector coupling is used as a free parameter which influences results at finite baryochemical potential. It is used to adjust the slope of the pseudocritical temperature of the chiral phase transition at low baryochemical potential and the scaled pressure shift accessible in lattice QCD simulations.
Our study, albeit presently performed at the meanfield level, supports the very existence of a critical point and favors its location within a region that is accessible in experiments at the NICA accelerator complex.
\PACS{
      {05.70.Jk}{Critical point phenomena}\and
      {11.10.Wx}{Finite-temperature field theory}\and
      {11.30.Rd}{Chiral symmetries}\and
      {12.38.Mh}{Quark-gluon plasma}\and
%      {12.39.Ki}{Relativistic quark model}\and
      {25.75.Nq}{Quark deconfinement, quark-gluon plasma production, and phase transitions}
      % \and
%     {PACS-key}{discribing text of that key}
     } % end of PACS codes
} %end of abstract
\authorrunning{{G. A. Contrera {\it et al.}}}
\titlerunning{Lattice QCD constrained CEP prediction in nonlocal PNJL models}
\maketitle

The search for the location of the critical endpoint (CEP) of first order
phase transitions in the QCD phase diagram is one of the objectives for
beam energy scan (BES) programs in relativistic heavy-ion collision
experiments at RHIC and SPS as well as in future ones at NICA and FAIR
which try to identify the parameters
of its position $(T_{\rm CEP}$, $\mu_{\rm CEP})$.
From a theoretical point of view, the situation is very blurry since the predictions for
this position form merely a skymap in the $T$-$\mu$ plane
\cite{Stephanov:2007fk}.

Lattice QCD results at zero and small
chemical potential $\mu$, show that the chiral and deconfinement transitions are crossover
with a pseudocritical temperature of $T_c(0)=154 \pm 9$ MeV
\cite{Bazavov:2011nk}.

However, at finite density lattice QCD suffers from the sign problem and only extrapolation or approximate techniques are available that work at finite quark densities.
Therefore, nonperturbative methods and effective models are inevitable tools in this region.
Up to now, such effective low-energy QCD approaches are not yet sufficiently developed to provide a unified approach to quark-hadron matter where hadrons appear as strongly correlated (bound) quark states that eventually dissolve into their quark (and gluon) constituents in the transition from the hadronic phase with confined quarks to the quark gluon plasma.
Since this transition shall be triggered by chiral symmetry restoration (by lowering the thresholds for hadron dissociation determined by in-medium quark masses), we expect that a first step towards a theoretical
approach to the QCD phase diagram is the determination of order parameters characterizing the QCD phases in a meanfield approximation for chiral quark models of different degree of sophistication.
As a consequence there appeared  a variety of possibilities for the structure of the QCD phase diagram and the position of the CEP in the literature. Let us mention few of them:
\begin{itemize}
\item no CEP at all \cite{Bratovic:2012qs}, with
crossover transition in the whole phase diagram,
\item no CEP, but a Lifshitz point \cite{Carignano:2010},
\item one CEP, but with largely differing predictions of its position
\cite{Stephanov:2007fk},
\item second CEP \cite{Kitazawa:2002bc,Blaschke:2003cv,Hatsuda:2006ps},
\item CEP and triple point, possibly coincident, considering another phase
(i.e. color superconducting \cite{Blaschke:2004cc} or quarkyonic
\cite{Andronic:2009gj} matter) at low temperatures and high densities.
\end{itemize}
This situation is far from being satisfactory in view of the upcoming experimental
programmes.
An exhaustive analysis should be performed to predict a CEP region as narrow as possible, considering only those effective models that
best reproduce recent lattice QCD results on the one hand and that obey constraints from heavy-ion collision experiments and compact star observations where available.

In the present contribution, we discuss the existence and location
of a CEP within the class of nonlocal chiral quark models coupled to the
Polyakov loop (PL) potential, with vector channel interactions, on the selfconsistent meanfield level, contrasting our results with those of the widely used local PNJL models that appear as limiting
case of the present approach.

The Lagrangian of these models is given by
\begin{eqnarray}
{\mathcal L} = \bar{q}(i\slashed{D}-m_0) q + {\mathcal L}_{\rm int} +
{\cal{U}}(\Phi)~,
\end{eqnarray}
where $q$ is the $N_{f}=2$ fermion doublet $q\equiv(u,d)^T$,
and $m_0$ is the current quark mass (we consider isospin symmetry, that is
$m_0=m_{u}=m_{d}$). The covariant derivative is defined as
$D_\mu\equiv \partial_\mu - iA_\mu$, where $A_\mu$ are color gauge fields.

The nonlocal interaction channels are given in the current-current coupling form by
 \begin{eqnarray}
{\mathcal L}_{\rm int}= -\frac{G_{S}}{2} \Big[ j_{a}(x)j_{a}(x)- j_{P}%
(x)j_{P}(x)\Big] {-}
\frac{G_V}{2} j_V(x)\, j_V(x),\nonumber\\
\end{eqnarray}
where the nonlocal generalizations of the currents are
\begin{eqnarray}
j_{a}(x)  &  =&\int d^{4}z\ g(z)\ \bar{q}\left(x+\frac{z}{2}\right)
\ \Gamma_{a}\ q\left(  x-\frac{z}{2}\right)  \ ,\nonumber\\
j_{P}(x)  &  =&\int d^{4}z\ f(z)\ \bar{q}\left(x+\frac{z}{2}\right)
\ \frac{i {\overleftrightarrow{\rlap/\partial}}}{2\ \kappa_{p}}
\ q\left( x-\frac{z}{2}\right) \ ,\nonumber\\
j_{V}(x)  &  =&\int d^4 z \ g(z)\ \bar{q}\left(x+\frac{z}{2}\right)
\,\gamma^0 \ q\left( x-\frac{z}{2}\right).
\label{eq:currents}
\end{eqnarray}
with $\Gamma_{a}=(\Gamma_{S},\Gamma_{P})=
(\leavevmode\hbox{\small1\kern-3.8pt\normalsize1},i\gamma_{5}\vec{\tau})$
for scalar and pseudoscalar currents respectively, and
$u(x^{\prime}){\overleftrightarrow{\partial}}v(x)=
u(x^{\prime})\partial_{x}v(x)-\partial_{x^{\prime}}u(x^{\prime})v(x)$.
The functions $g(z)$ and $f(z)$ in Eq.~(\ref{eq:currents}) are
nonlocal covariant form factors characterizing the corresponding
interactions.
The scalar-isoscalar component of the $j_{a}(x)$
current will generate the momentum dependent quark mass in the
quark propagator, while the ``momentum'' current, $j_{P}(x),$ will
be responsible for a momentum dependent wave function
renormalization (WFR) of this propagator.
Note that the relative strength between both interaction terms is controlled
by the mass parameter $\kappa_{p}$ introduced in Eq.~(\ref{eq:currents}).
Finally, $j_{V}(x)$ represents the vector channel interaction current, whose
coupling constant $G_V$ is usually taken as a free parameter.

In what follows it is convenient to Fourier transform into momentum space.
Since we are interested in studying the characteristics of the chiral phase
transition we have to extend the effective action to finite temperature $T$
and chemical potential $\mu$.
In the present work this is done by using the Matsubara imaginary time formalism.
Concerning the gluon degrees of freedom we employ the PL
extension of nonlocal chiral quark models according to previous works
\cite{Contrera:2007wu,Contrera:2010kz,Horvatic:2010md,Radzhabov:2010dd,Benic:2013eqa}, i.e. assuming that the quarks move in a background color gauge field $\phi = iA_0 = i g\,\delta_{\mu 0}\, G^\mu_a \lambda^a/2$, where $G^\mu_a$ are
the SU(3) color gauge fields and $\lambda^a$ are the Gell-Mann matrices. Then
the traced PL $\Phi$, which is taken as order parameter of confinement, is given by
$\Phi=\frac{1}{3} {\rm Tr}\, \exp( i \phi/T)$. Then, working in the so-called Polyakov gauge, in which the
matrix $\phi$ is given by a diagonal representation $\phi = \phi_3 \lambda_3 + \phi_8 \lambda_8$. At vanishing chemical potential, owing to the charge conjugation properties of the QCD Lagrangian, the traced PL is expected to be a real quantity.
Since $\phi_3$ and $\phi_8$ have to be real-valued~\cite{Rossner:2007ik}, this condition implies
$\phi_8 = 0$.
In general, this need not be the case at finite $\mu$ \cite{Dumitru:2005ng,Fukushima:2006uv}.
As, e.g., in Refs.~\cite{Contrera:2010kz,Rossner:2007ik,Abuki:2008ht,GomezDumm:2008sk} we will assume that the potential $\cal U$ is such that the condition $\phi_8=0$ is well satisfied for the range of values of $\mu$ and $T$ investigated here.
The mean field traced PL is then given by
\begin{equation}
\Phi = \frac{1}{3} \Big[ 1 + 2\,\cos (\phi_3/T) \Big].
\label{eq:Phi}
\end{equation}

In the present work we have chosen a $\mu$-dependent logarithmic effective potential described in
\cite{Dexheimer:2009va}.
\begin{eqnarray}
{\cal{U}}(\Phi,T,\mu)=&&(a_0T^4+a_1\mu^4+a_2T^2\mu^2)\Phi^2 \nonumber\\
&& + a_3T_0^4\ln{(1-6\Phi^2+8\Phi^3-3\Phi^4)},
\label{eq:PL_pot}
\end{eqnarray}
where the parameters are $a_0=-1.85$, $a_1=-1.44$x$10^{-3}$, $a_2=-0.08$,
$a_3=-0.40$.
For the $T_0$ parameter we use the value corresponding to two flavors $T_0= 208$ MeV, as has been suggested in
Ref.~\cite{Schaefer:2007pw} and already employed in nonlocal PNJL models in \cite{Horvatic:2010md,Pagura:2012}.

Finally, in order to fully specify the nonlocal model under consideration we
set the model parameters as well as the form factors $g(q)$ and $f(q)$
following Refs.~\cite{Contrera:2010kz}, \cite{Noguera:2008prd78} and \cite{Contrera:2012wj}, i.e. considering two different types of functional dependencies for these form
factors: exponential forms
%(Set A and Set B)

\begin{eqnarray}
(\textrm{Set A}) \quad \left\{
\begin{array}{l}
g(p)= \mbox{exp}\left(-p^{2}/\Lambda_{0}^{2}\right) \\
f(p)=0
\end{array}
\right.~,
\label{eq:setA}
\end{eqnarray}
\begin{eqnarray}
(\textrm{Set B}) \quad \left\{
\begin{array}{l}
g(p)= \mbox{exp}\left(-p^{2}/\Lambda_{0}^{2}\right) \\
f(p)= \mbox{exp}\left(-p^{2}/\Lambda_{1}^{2}\right)
\end{array}
\right. ~,
\label{eq:setB}
\end{eqnarray}
and Lorentzians with WFR
\begin{eqnarray}
(\textrm{Set C}) \quad \left\{
\begin{array}{l}
g(p)  = \frac{1+\alpha_z}{1+\alpha_z\ f_z(p)} \frac{\alpha_m \ f_m(p) -m\ \alpha_z f_z(p)}
{\alpha_m - m \ \alpha_z } \\
f(p)  = \frac{ 1+ \alpha_z}{1+\alpha_z \ f_z(p)} f_z(p)\
\end{array}
\right.~,
\label{eq:setC}
\end{eqnarray}
where
\begin{eqnarray}
f_{m}(p) = \left[ 1+ \left( p^{2}/\Lambda_{0}^{2}\right)^{3/2} \right]^{-1}, \nonumber\\
f_{z}(p) = \left[ 1+ \left( p^{2}/\Lambda_{1}^{2}\right) \right]^{-5/2},
\label{eq:setC2}
\end{eqnarray}
and $\alpha_m = 309$ MeV, $\alpha_{z}=-0.3$.
Further details on the parameters can be found in Ref.~\cite{Contrera:2012wj} and references quoted therein.

Within this framework the thermodynamic potential in the mean field approximation (MFA) reads
\begin{eqnarray}
\Omega^{\rm MFA} = && - \,{4 T} \sum_{c} \sum_n \int \frac{d^3\vec p}{(2\pi)^3} \, \,
\mbox{ln} \left[ \frac{ (\tilde{\rho}_{n, \vec{p}}^c)^2 + M^2(\rho_{n,\vec{p}}^c)}{Z^2(\rho_{n, \vec{p}}^c)}\right]  \nonumber\\
&& + \frac{\sigma_1^2 + \kappa_p^2\ \sigma_2^2}{2\,G_S} - \frac{\omega^2}{2 G_V}\ + {\cal{U}}(\Phi ,T).
\label{granp}
\end{eqnarray}
where $M(p)$ and $Z(p)$ are given by
\begin{eqnarray}
M(p) & = & Z(p) \left[m + \sigma_1 \ g(p) \right] ,\nonumber\\
Z(p) & = & \left[ 1 - \sigma_2 \ f(p) \right]^{-1}.
\label{mz}
\end{eqnarray}

In addition, we have defined \cite{Blaschke:2007ri}
\begin{equation}
\Big({\rho_{n,\vec{p}}^c} \Big)^2 =
\Big[ (2 n +1 )\pi  T - i \mu + \phi_c \Big]^2 + {\vec{p}}\ \! ^2 \ ,
\label{eq:rho}
\end{equation}
where the quantities $\phi_c$ are given by the relation $\phi_c =
{\rm diag}(\phi_r,\phi_g,\phi_b)$. Namely, $\phi_c = c \ \phi_3$
with $c = 1,-1,0$ for $r,g,b$, respectively.

In the case of $\Big({\tilde{\rho}_{n,\vec{p}}^c}\Big)$ we have used the same definition as in Eq.(\ref{eq:rho}) but shifting the
chemical potential
according to \cite{Contrera:2012wj}
\begin{equation}
\tilde{\mu} = \mu \; - \omega \; g(p) \; Z(p).
\label{mutilde}
\end{equation}

We also want to include in our analysis the results arising from a local PNJL model based on \cite{Fukushima2008} with two
flavors instead of three.
Moreover, we consider that the chemical potential is shifted by

\begin{equation}
\tilde{\mu} = \mu \; - \omega.
\label{mutilde0}
\end{equation}

\begin{figure}[hbtp]
\begin{center}
\includegraphics[width=1.2 \linewidth]{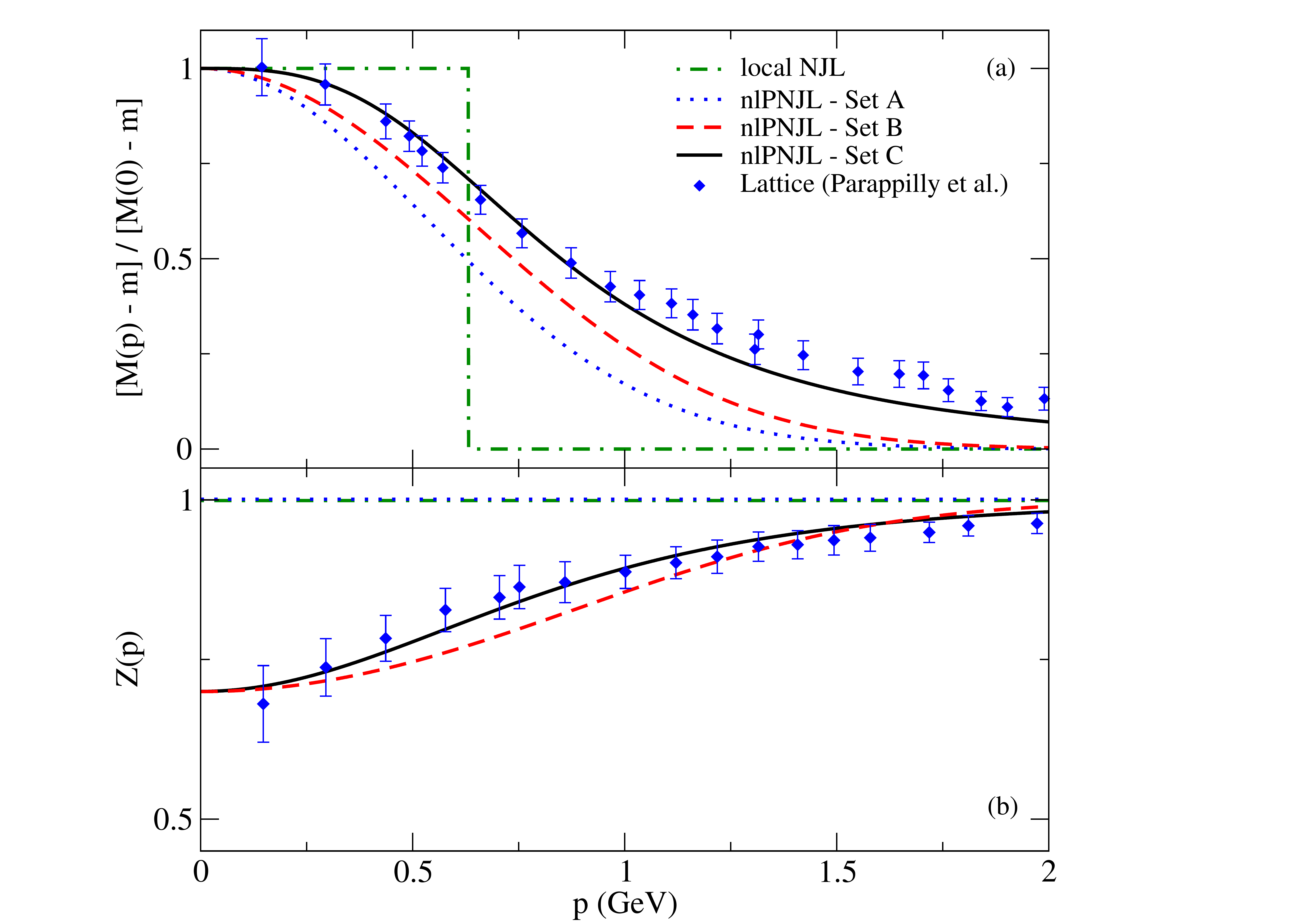}  %{WFR_lattice_fit.pdf}
\end{center}
\caption{Normalized dynamical masses for the different form factors under study and wave function renormalization for Set B and Set C, fitted to lattice QCD data \cite{Parappilly:2005ei}. (Adapted from Ref.~\cite{Noguera:2008prd78}). }
\label{fig:1}
\end{figure}

$\Omega^{\rm MFA}$ turns out to be divergent and, thus, needs to be
regularized. For this purpose we use the same prescription as in
Refs.~\cite{Dumm:2005,Contrera:2010kz}.
The mean field values $\sigma_{1,2}$, $\omega$ and $\phi_3$ at a given
temperature or chemical potential, are obtained from a set of four coupled
``gap'' equations which come from the minimization of the regularized
thermodynamic potential, that is
\begin{equation}
\frac{\partial\Omega^{\rm MFA}_{\rm reg}}{\partial\sigma_{1}} =
\frac{\partial\Omega^{\rm MFA}_{\rm reg}}{\partial\sigma_{2}} =
\frac{\partial\Omega^{\rm MFA}_{\rm reg}}{\partial\omega} =
\frac{\partial\Omega^{\rm MFA}_{\rm reg}}{\partial\phi_3} = 0.
\label{fullgeq}
\end{equation}
%%%%%%%%%%%%%%%%%%%%%%%%%%%%%%%%%%%%%%%%%%%%%%%%%%%%%%%%%%%%%%%%%%%%%%%%
As a starting point we consider the local NJL, using in this case the parameters as in \cite{GomezDumm:2008sk}.
This resembles the local limit of the present model and results can be compared, e.g., with
Refs.~\cite{Bratovic:2012qs} and \cite{Friesen:2014mha}.
As in our previous work \cite{Contrera:2012wj}, the model inputs have been constrained with results from lattice QCD studies.
In particular, the form factors of the nonlocal interaction can be chosen such as to reproduce the dynamical mass function $M(p)$
and the WFR function $Z(p)$ of the quark propagator in the vacuum \cite{Parappilly:2005ei}.
In Fig.~\ref{fig:1} we show the shapes of normalized dynamical masses and WFR for the models under discussion here, i.e., the nonlocal models of Set A (rank-one), Set B and Set C (rank-two) as well as the local limit.
From this figure we see that Set C fits best the normalized dynamical mass.

%%%%%%%%%%%%%%%
%Figure 1
\begin{figure}[hbtp]
\begin{center}
\includegraphics[width=1.0 \linewidth]{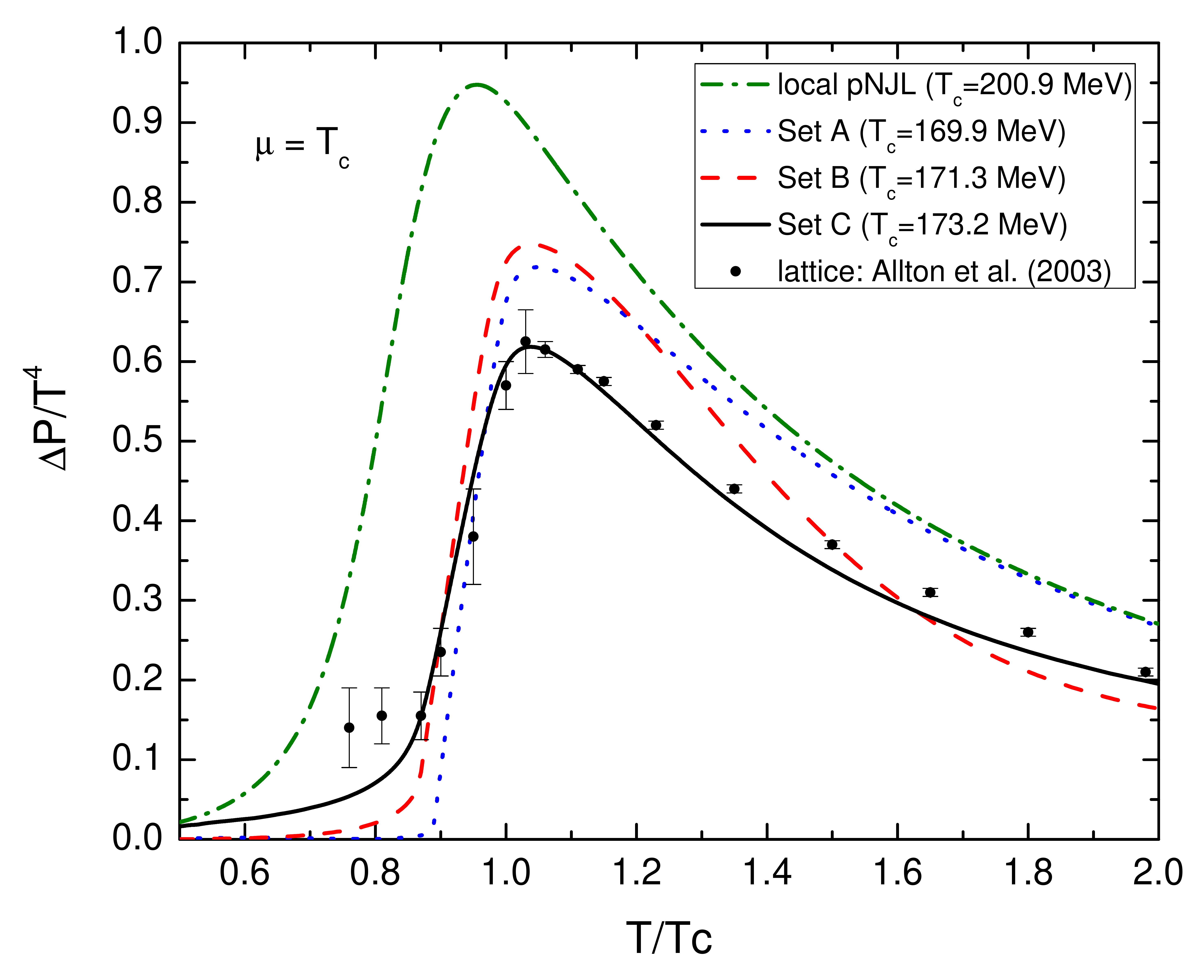}
\includegraphics[width=1.45 \linewidth]{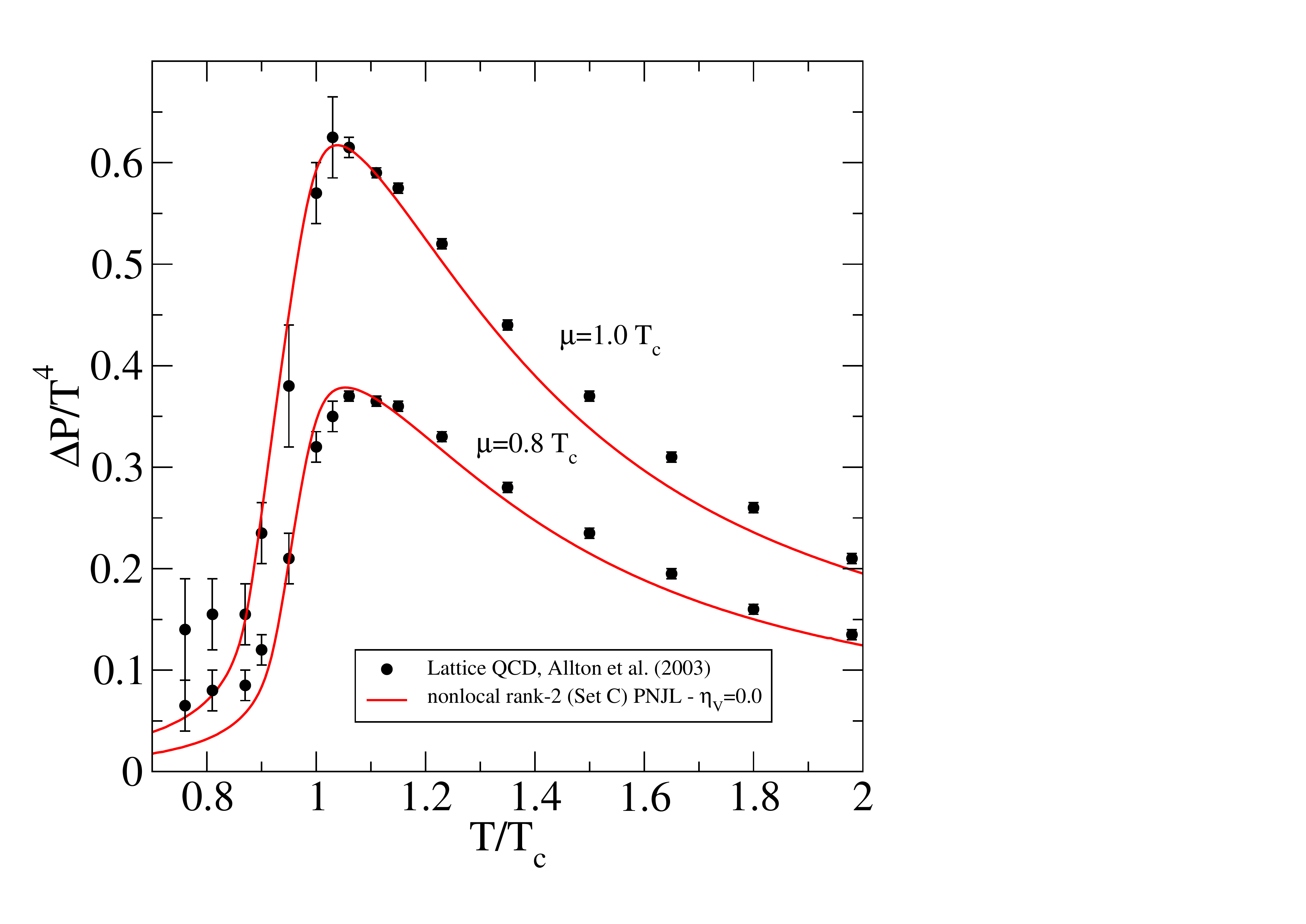}
\end{center}
\caption{
The scaled pressure shift $\Delta P/T^4$ as a function of the scaled temperature $T/T_c$, for $G_V=0.0$.
Upper panel: Comparison of the lattice QCD data \cite{Allton:2003vx} with results of local and nonlocal pNJL models.
Lower panel: Comparison of Set C results with lattice QCD \cite{Allton:2003vx} for two chemical potentials $\mu=1.0~T_c$ and $\mu=0.8~T_c$.}
\label{fig:2}
\end{figure}

%%%%%%%%%%%%%%%
%
Now we are in the position to discuss the results for the thermodynamics of the nonlocal PNJL models, starting from the pressure $P(\mu,T)=-\Omega^{\rm MFA}_{\rm reg}$.
In the upper panel of  Fig.~\ref{fig:2} we compare the pressure shift
$\Delta P= P(\mu, T) - P(0, T )$ scaled with $T^4$ as a function of $T/T_c$ for all sets of parametrizations shown in  Fig.~\ref{fig:1} to lattice results from \cite{Allton:2003vx,Fodor:2002km}.
It is obvious that the best obtained fit corresponds to Set C, supporting the robustness of rank-2 Lorentzian parametrization.
In the lower panel of Fig.~\ref{fig:2} we show a comparison between Set C and lattice results \cite{Allton:2003vx}, for $G_V=0.0$ (i.e. without vector interactions).

%%%%%%%%%%%%%%%%%%
%Figure 2
\begin{figure}[hbtp]
\begin{center}
\includegraphics[width=0.95 \linewidth]{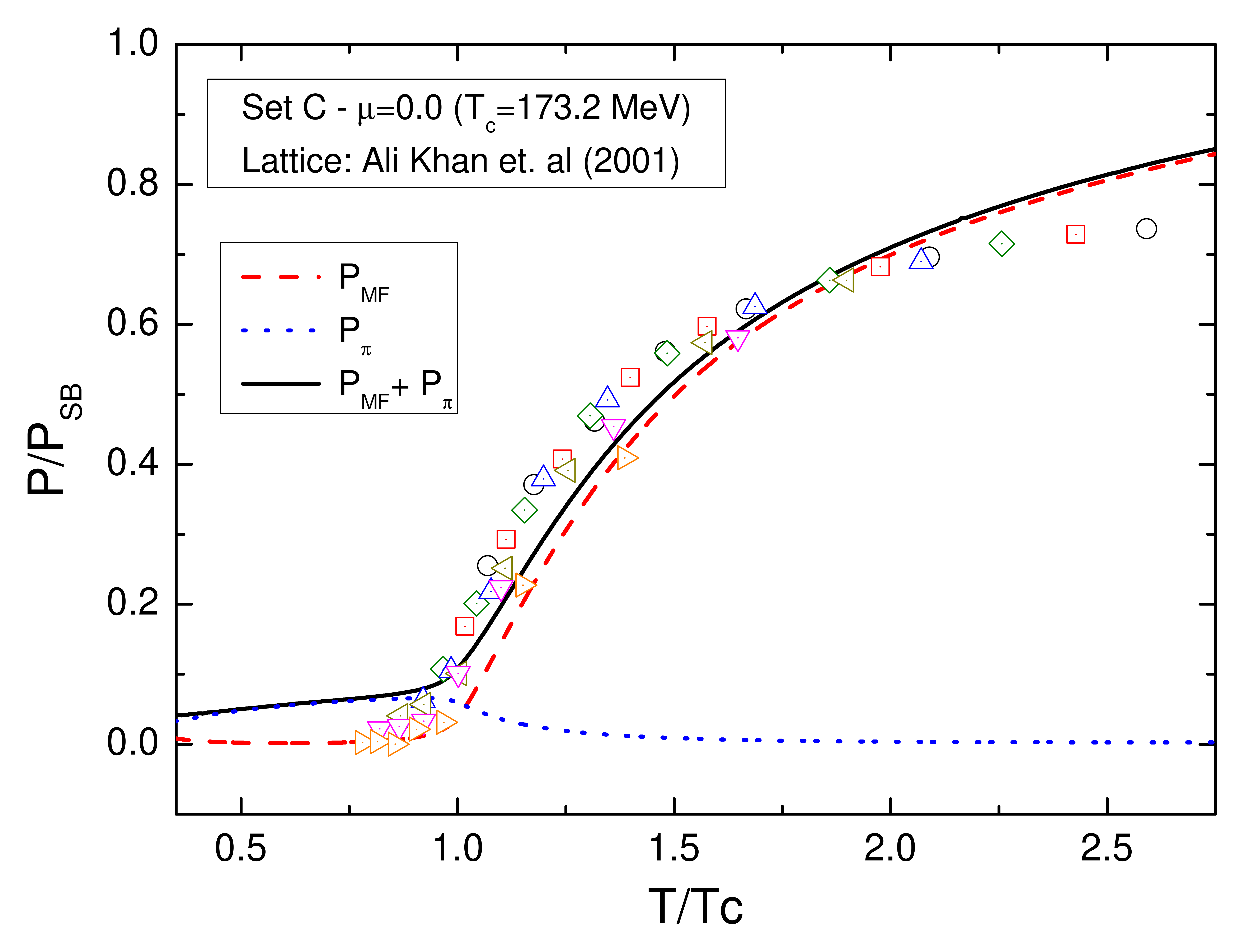}  %{Pressure with pions_lattice_fit.pdf}
\end{center}
\caption{Comparison of the pressure in units of its Stefan-Boltzmann value between Set C
(with and without pion pressure) and lattice QCD results \cite{AliKhan:2001ek} as function of $T/T_c$.}
\label{fig:3}
\end{figure}

%%%%%%%%%%%%%%%%%
In Fig.~\ref{fig:3} we show a comparison of the pressure normalized to the Stefan-Boltzmann limit
($P_{\rm SB}$) to lattice QCD results from \cite{AliKhan:2001ek}.
In this figure it can be seen that the pion pressure dominates the pressure dependence for $T<T_c$
and quickly vanishes for $T>T_c$ as a result of the Mott dissociation of the pion.
%%%%%%%%%%%%%%%%
%Figure 3

%%%%%%%%%%%%%%%%%

From the above results, Set C appears to be the appropriate choice to best fit lattice QCD results.

\begin{figure}[hbtp]
\begin{center}
\includegraphics[width=0.95 \linewidth]{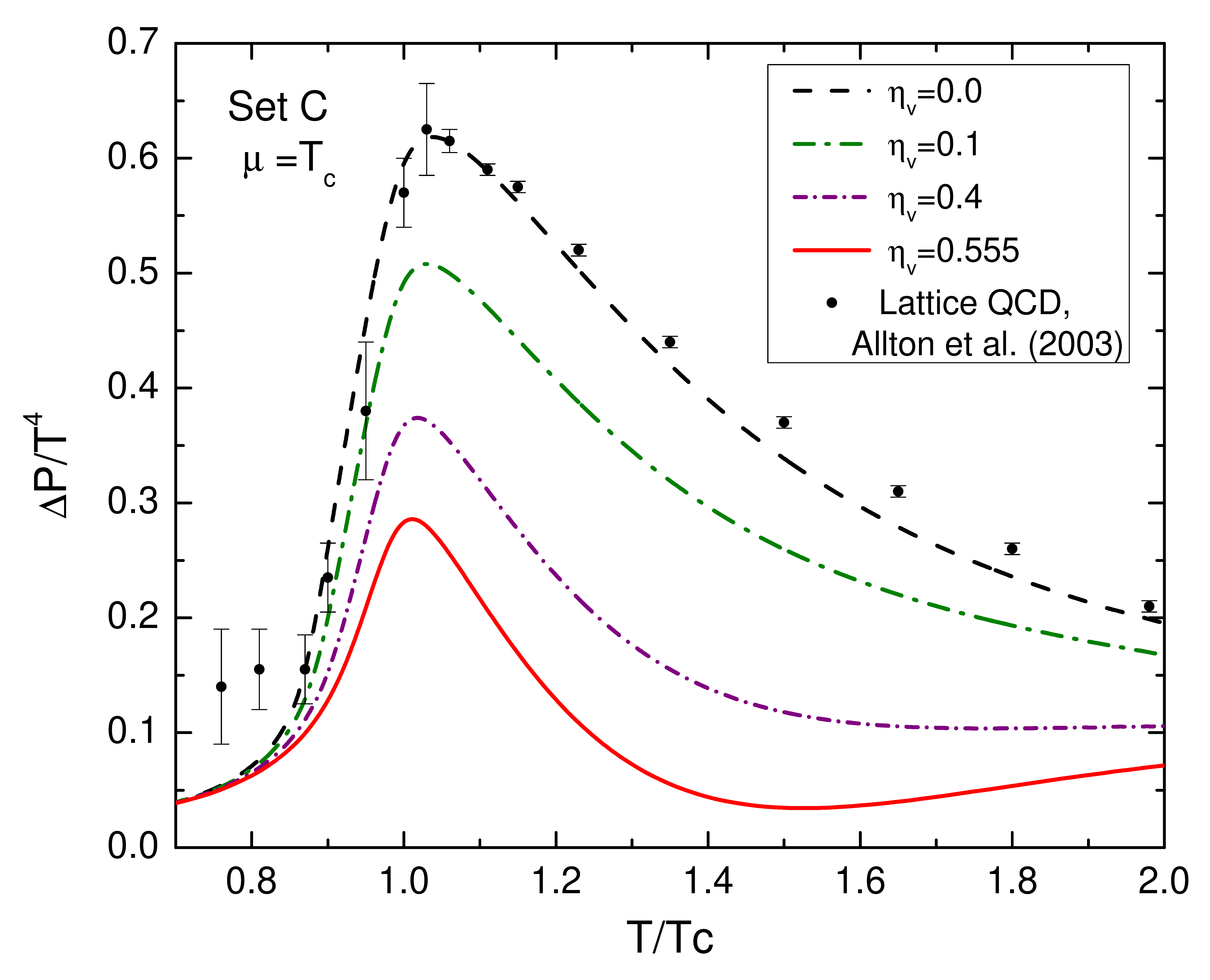}
\end{center}
\caption{Comparison between Set C  results for different
values of the vector coupling parameter $\eta_V$ and lattice QCD data \cite{Allton:2003vx}.}
\label{fig:4}
\end{figure}

As mentioned above, the vector coupling $G_V$ is considered a free parameter.
In Fig.~\ref{fig:4} we show again $\Delta P/T^4$ vs. $T/T_c$ as in Fig.~\ref{fig:2}, but now considering different vector coupling parameters $\eta_V = G_V/G_S$ just for Set C and $\mu=T_c$.
The best agreement with lattice QCD data is obtained for $\eta_V = 0$, i.e. for the deactivated vector channel.
The consequences will be discussed below.

%%%%%%%%%%%%%%%%
%Figure 4

%%%%%%%%%%%%%%%%%

In MFA the vector coupling channel has a direct influence on the $\mu-$dependence of the pseudocritical temperature  $T_c(\mu)$ in the QCD phase diagram.
In lattice QCD this dependence has been analyzed by Taylor expansion techniques \cite{Kaczmarek:2011zz} as
\begin{equation}
T_c(\mu)/T_c(0)= 1 - \kappa (\mu/T)^2 + {\mathcal{O}}[(\mu/T)^4],
\label{eq:lattice_cross}
\end{equation}
with $\kappa=0.059 (2) (4)$ being the curvature.
We will refer to this situation as LR I (lattice results I).
In the same way, the vector coupling channel can be tuned to reproduce the curvature given by recent lattice QCD results based on imaginary chemical potential technique which gives $\kappa=0.1341 \pm 0.019$ \cite{Bellwied:2015rza} and $\kappa=0.1215 \pm 0.018$ \cite{Bonati:2015bha}, we call it LR II.

The curvatures can be determined from the phase diagrams. To do so, we plotted the pseudocritical temperatures of the crossover
transitions as a function of $(\mu/T)^2$ for different values of the vector coupling parameter $\eta_V$.
Then, the curvatures can be obtained from the
slope of the straight lines in the region of low $(\mu/T)$ values.

\begin{figure}[hbtp]
\centering
\includegraphics[width=1.0 \linewidth]{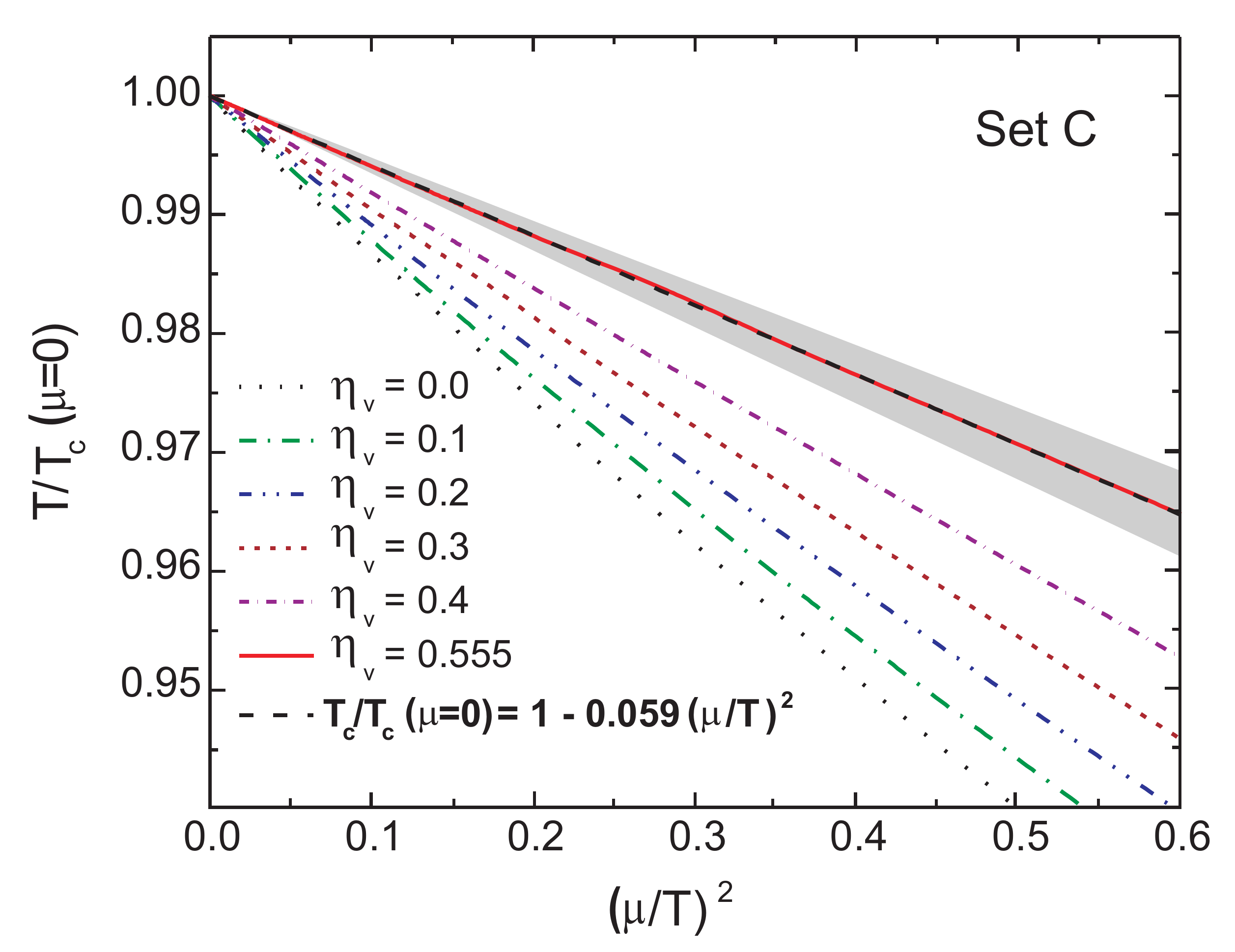}
\caption{(Color online) Chiral crossover transitions at low values of $\mu /T$ for different values of the vector coupling ratios
$\eta_V= G_V/G_S$ for Set C. The dashed line corresponds to the lattice QCD prediction of $\kappa=0.059 (2) (4)$, LR I, \cite{Kaczmarek:2011zz}.}
\label{fig:5}
\end{figure}

An example of this is shown in Figure \ref{fig:5} for set C (the corresponding plots for the other sets are qualitatively very similar).
The fit (\ref{eq:lattice_cross}) of the lattice QCD results, {LR I}, is also shown.
The grey zone corresponds to the error in the coefficient $\kappa$ obtained in \cite{Kaczmarek:2011zz}.

%%%%%%%%%%%%%%%%%%%%%%%%%%%%%%%%%%%%%%%%%%%%%%%%%%%%%
%Figure 5

%%%%%%%%%%%%%%%%%%%%%%%%%%%%%%%%%%%%%%%%%%%%%%%%%%%%%

\begin{figure}[hbtp]
\centering
\includegraphics[width=1.10 \linewidth]{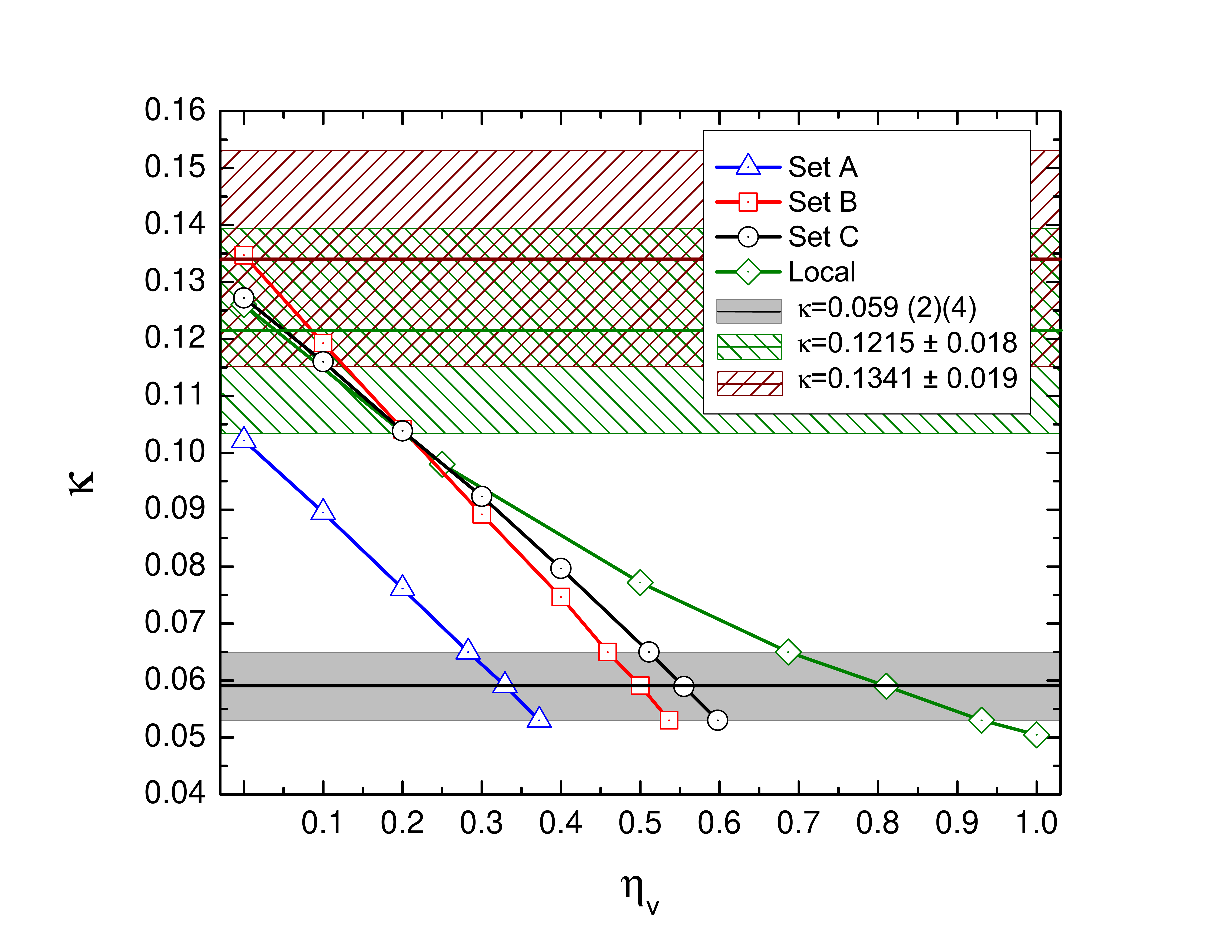}
\caption{(Color online) Curvature $\kappa$ of the pseudocritical temperature $T_c(\mu)$ of the chiral crossover transition at low
values of $\mu /T$.
The black line corresponds to the lattice QCD prediction of
$\kappa=0.059 (2) (4)$
\cite{Kaczmarek:2011zz}, while the brown (green) lines with hatched error regions come from analyses
within the imaginary chemical potential method, resulting in
$\kappa=0.1341\pm 0.019$
\cite{Bellwied:2015rza}
($\kappa=0.1215 \pm0.018$
\cite{Bonati:2015bha}).
}
\label{fig:6}
\end{figure}

In Fig.~\ref{fig:6} we compare the lattice QCD results for both, LR I and LR II, with the values for the coefficient $\kappa$ obtained within all PNJL models under study, considering different values of $\eta_V$.

%%%%%%%%%%%%%%%%%%%%%%%%%%%%%%%%%%%%%%%%%%%%%%%%%%%%%
%Figure 6

Note that for fitting the lattice QCD value in the local model a larger vector
coupling is required than in the nonlocal ones, as is shown in Ref.\cite{Contrera:2012wj}.
Also the absolute value of the critical temperature $T_c(0)$ in the local
model is significantly different (larger) than in the nonlocal one.

\begin{figure}[hbtp]
\centering
\includegraphics[width=1.0\linewidth]{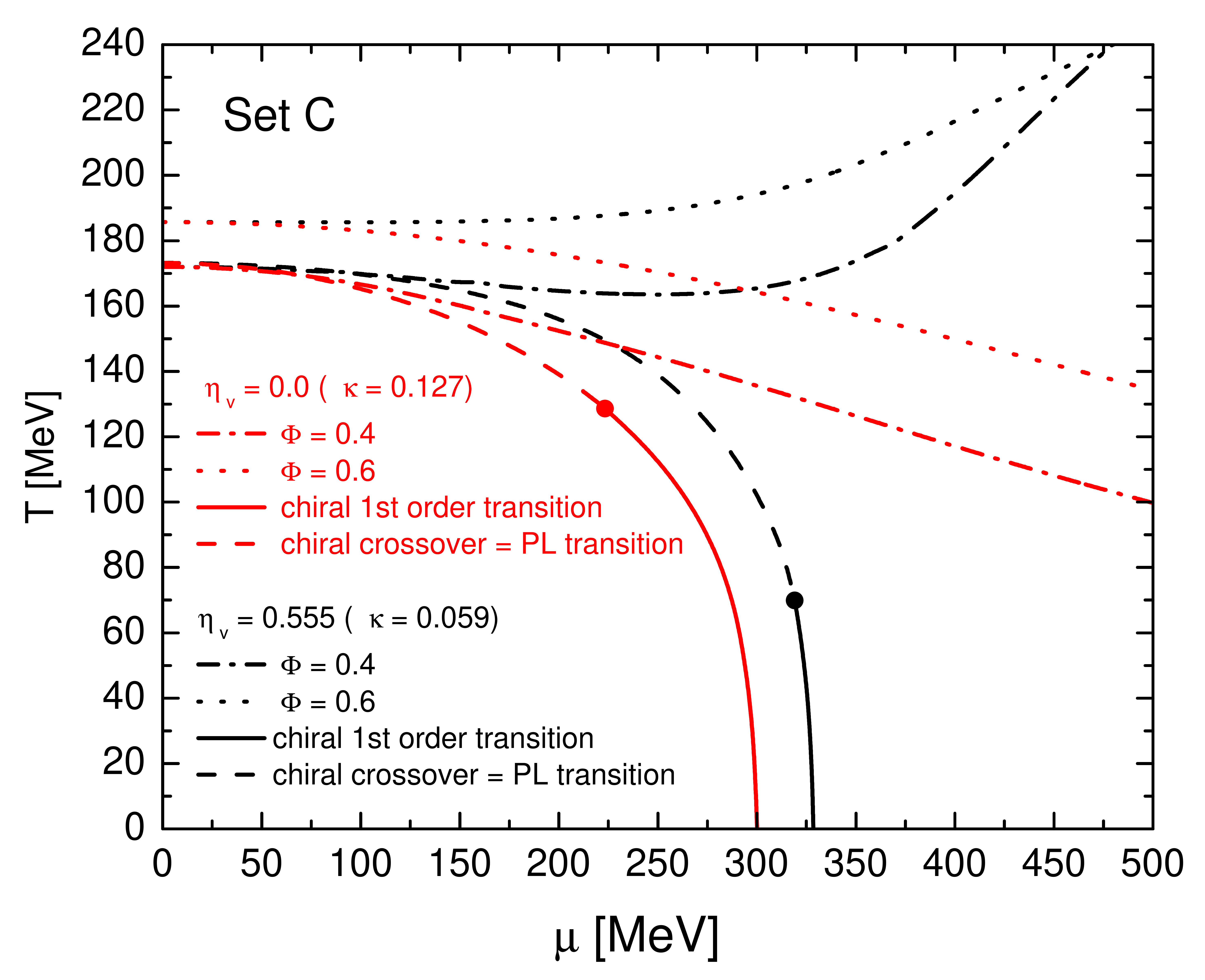}   %{Ph_Diag_U3_T0_208.pdf}
\caption{Phase diagrams with (pseudo)critical temperatures $T_c(\mu)$ and critical endpoints
for set C parametrization of the nonlocal rank-2 PNJL model for the two cases $\eta_v=0.0$ and
$\eta_v=0.555$.
Dashed (full) lines correspond to the chiral crossover (first order) transitions.
The PL transition lines, defined by the peak of the PL susceptibility, coincide with the chiral crossover.
The dash-dotted and dotted lines represent the deconfinement transition range, i.e. $\Phi=0.4$ and $\Phi=0.6$, respectively.
}
\label{fig:7}
\end{figure}

The phase diagram with (pseudo-)critical temperatures $T_c(\mu)$ and critical points for Set C, i.e. the parametrization that best fits lattice QCD results (see Figs.~\ref{fig:1} to \ref{fig:3}) are shown in
Fig.~\ref{fig:7}.
The finite vector coupling parameter $\eta_V$ is chosen to reproduce the curvature value {for LR I} at
low $\mu$.
We found that switching off the vector interaction can reproduce both, the curvature obtained in LR II and the $\Delta P/T^4$ vs. $T/T_c$ from Fig.~\ref{fig:2}
\footnote {Note that in Refs. \cite{Bellwied:2015rza,Bonati:2015bha} the values are quoted in the $T$-$\mu_B^2$ plane, while we are
addressing the quark chemical potential instead of the baryon one. So the difference in the $\kappa$ coefficients is a factor
9.}.

Now we also considered the pseudo-critical deconfinement lines for $\Phi=0.4 - 0.6$ in Fig.~\ref{fig:7}.
To determine them we proceed as follows: we set the fixed value of $\Phi$ in Eq.~(\ref{eq:Phi}), obtaining a relation between $\phi_3$ and T.
Then we replace this relation in the thermodynamic potential and minimize it respect to the mean fields by solving the Eqs.~(\ref{fullgeq}).
Thus, for each value of $\Phi$ we have the corresponding values of $\phi_3$ and T that satisfy the Eq.~(\ref{eq:Phi}) and the gaps equations for the desired value of $\Phi$.
Those are the deconfinement lines at fixed $\Phi$ shown in Fig.~\ref{fig:7}. However, this tells us only that the PL changes quickly with $T$, but not how much confining the model is still at high $T$.
Since obviously the change occurs at absolute values of $\Phi \le 0.4$, there are still strong
color correlations present in the system at high temperatures.

In order to estimate the region in the phase diagram where we expect color correlations as measured by the value of the Polyakov loop to be strong, we choose to show the lines of constant $\Phi$.
Interestingly, we find that in the presence of a vector meanfield the approach to the free quasiparticle case ($\Phi=1$) is inhibited.
It is clear that this is the more so the larger the chemical potential is since the vector meanfield is proportional to the baryon density which increases with $\mu$.

On the other hand, the PL transition as defined by the peak of the PL thermal susceptibility coincides
nicely with that of the crossover chiral transition.

%%%%%%%%%%%%%%%%%%%%%%%%%%%%%%%%%%%%%%%%%%%%%%%%%%%%%

%Figure 7

The values for $T_c(\mu=0)$ (in MeV) are 169.9, 171.3, 173.2 and 200.9, for Sets A, B, C and local, respectively. These results
indicate that the $T_c(0)$ of nonlocal
covariant PNJL models is rather insensitive to the choice of the form factors
parametrizing the momentum dependence (running) of the dynamical mass function and
the WFR function of the quark propagator as measured on the lattice at zero temperature
\cite{Parappilly:2005ei}, whereas the position of the CEP and critical chemical
potential at $T=0$ strongly depends on it.

In view of this finding, the absence of a CEP reported in the local limit \cite{Bratovic:2012qs} as well as the result for Set A without WFR seems to be less realistic (see Ref.\cite{Contrera:2012wj}).

On the other hand, more recent lattice results \cite{Bellwied:2015rza,Bonati:2015bha} (LR II) suggest higher values for the $\kappa$ coefficient, which implies a lower or even vanishing vector coupling
parameter $\eta_V$, giving better agreement for $\Delta P/T^4$ vs. $T/T_c$.

Considering the different lattice constraints for $\kappa$ and that the best fitting rank-2 nonlocal PNJL model is the one with Lorentzian form factors (Set C parametrization), the region in the QCD phase diagram where the CEP should be located according to our study, would be determined as
$(T_{\rm CEP},\mu_{\rm CEP})=(128.6~{\rm MeV}, 223.3~{\rm MeV})$ for $\eta_V=0.0$ and
$(T_{\rm CEP},\mu_{\rm CEP})=(69.9~{\rm MeV},319.1~{\rm MeV})$ for $\eta_V=0.555$.
This region is highlighted in Fig.~\ref{fig:7}.

With this result of the present study we arrive at our conclusion for the NICA experiments which are devoted to the study of the quark-hadron mixed phase that shall be located at $T<T_{\rm CEP}$ and
$\mu>\mu_{\rm CEP}$.
We shall estimate whether the planned energy ranges of the BM@N experiment
($E_{\rm lab}=2-4$ A GeV) and of the MPD experiment ($\sqrt{s_{NN}}=4 - 11$ GeV)
are suitable for accessing the region of the mixed phase that follows from our study.
To this end we use a parametrization for the chemical freeze-out temperature in the QCD phase diagram by Andronic et al. \cite{Andronic:2005yp}
\begin{equation}
T_{\rm freeze} = T_{\rm lim} \left(1-\frac{1}{0.7+[\exp(\sqrt{s_{NN}}({\rm GeV}))-2.9]/1.5} \right)
\end{equation}
as obtained from statistical model analyses of hadron production in heavy-ion collision experiments,  with
$T_{\rm lim}=161\pm 4 $ MeV.
According to this parametrization and to the LR-I motivated nonlocal PNJL model with vanishing vector
coupling, one should expect signals of a first order phase transition in collisions with
$\sqrt{s_{NN}}\lsim 6$ GeV, right in the middle of the MPD energy scan.
For the LR-II motivated parametrization with $\eta_V=0.555$ the CEP is at such a low temperature that
only the BM@N experiment has a chance to access the mixed phase, at laboratory energies
$E_{\rm lab}\lsim 3$ A GeV, within the range of this fixed target experiment but too low for the
collider experiment MPD.

The main conclusion of this study is that for the search of CEP signatures and
the investigation of properties of the quark-hadron mixed phase in BES programmes
the energy range of the NICA and FAIR facilities shall be particularly promising.
We find a certain preference for a CEP position at a critical temperature
$T_{\rm CEP}\sim 130$ MeV, expected to be crossed in collisions with $\sqrt{s_{NN}}\sim 6~$GeV,
corresponding to $E_{\rm lab}\sim 18~$A~GeV.
This energy could not be reached from above by the present RHIC BES programme of the
STAR experiment (lower limit $\sqrt{s_{NN}}= 7.7~$GeV and also not in the NA49 experiment
at CERN SPS (lower limit $E_{\rm lab}=20~$A~GeV).
Just the ongoing NA61-SHINE experiment has a lower limit of $E_{\rm lab}=13~$A~GeV in their energy scan that would allow to cross this suspected critical point (unless it would be located at slightly lower temperature).
The old AGS experiment at BNL did not quite reach this energy (maximum energy $E_{\rm lab}=10.74~$A~GeV) and were also not designed for a CEP search, while the energies at the GSI SIS are probably too low to reach the phase transition border (maximum energy $E_{\rm lab}=2~$A~GeV).
The theoretical investigation of the baryon stopping as  a probe of the onset of deconfinement in
heavy-ion collisions \cite{Ivanov:2012bh} has revealed a characteristic "wiggle" structure in the curvature of the proton rapidity distribution at $\sqrt{s_{NN}}\sim 6~$GeV which remains robust also when applying
acceptance cuts of the MPD experiment \cite{Ivanov:2015vna} at NICA.
The MPD experiment at NICA will be the first dedicated heavy-ion collision experiment of the third generation to fully cover the suspected location of the CEP and thus to enter the mixed phase region via a first order phase transition.
Contrary to studies with the local NJL model which first in \cite{Sasaki:2006ws}
and more recently in \cite{Bratovic:2012qs} have shown that a CEP may be absent
at all in the phase diagram for finite coupling in the repulsive vector channel, the present work with nonlocal PNJL models constrained by lattice QCD propagator data has shown that the
presence of a CEP in the QCD phase diagram is rather robust against even stronger
vector channel interactions.

Finally, we would like to mention currently ongoing developments of our approach to the QCD phase diagram
and the EoS of matter under extreme conditions:
\begin{itemize}
\item Extension of the model to 2+1 flavours in order to properly compare with modern lattice QCD results
\cite{Borsanyi:2013bia,Bazavov:2014pvz}.
\item Investigation of the robustness of the results
of the nonlocal PNJL models when modifying the choice of the Polyakov-loop
potential taking into account recent developments
\cite{Sasaki:2012bi,Ruggieri:2012ny,Fukushima:2012qa}.
\item Addition of quark pair interaction channels and the possibility of color superconducting quark matter phases
\cite{GomezDumm:2008sk,Blaschke:2010ka}
\item Going beyond the mean field approximation within the Beth-Uhlenbeck approach \cite{Blaschke:2013zaa},
where hadronic correlations of quarks are included with their spectral weight showing both, bound and scattering parts, with a characteristic medium dependence that exhibits the Mott transition
(dissociation of hadronic bound states into the scattering continuum at finite temperatures and chemical potentials).
\end{itemize}

A key quantity for such studies are is hadronic phase shifts.
%spectral function and
First results using a generic ansatz \cite{Blaschke:2015nma}
%\cite{Blaschke:2011ry}
for joining the hadron resonance gas and PNJL approaches are promising
%\cite{Turko:2011gw}.
\cite{Dubinin:2015glr}.
The inclusion of baryonic correlations is possible and has been started at finite temperatures
\cite{Blaschke:2015sla} and will be extended to the full phase diagram.
As the behavior of the hadronic spectral functions and in particular the location of the Mott transitions
in the QCD phase diagram is essentially determined by the dynamical quark mass functions, a key
issue for future research is to investigate the backreaction of hadronic correlations on these order parameters.
We expect that by following the Beth-Uhlenbeck approach one will have a methodic basis for answering   the question about the robustness of predictions for the structure of the QCD phase diagram at the meanfield level discussed here.

\subsection*{Acknowledgement}
We would like to thank to O. Kaczmarek for useful comments and discussions.
D.B.\ acknowledges hospitality and support during his visit at University of
Bielefeld and funding of his research provided by the Polish NCN within
the ``Maestro'' grant programme, under contract number UMO-2011/02/A/ST2/00306.
%No.\ NN 202 231837 and
%as well as by the Russian Fund for Basic Research under Grant No.\ 11-02-01538-a.
G.C. and A.G.G. acknowledge financial support of CONICET and UNLP (Argentina).


\begin{thebibliography}{99}

%\cite{Stephanov:2007fk}
\bibitem{Stephanov:2007fk}
  M.~A.~Stephanov,
  %``QCD phase diagram: An Overview,''
  PoS LAT {\bf 2006}, 024 (2006).
%  [hep-lat/0701002].
  %%CITATION = HEP-LAT/0701002;%%

%\cite{Bazavov:2011nk}
\bibitem{Bazavov:2011nk}
  A.~Bazavov, T.~Bhattacharya, M.~Cheng, C.~DeTar, H.~T.~Ding, S.~Gottlieb, R.~Gupta and P.~Hegde {\it et al.},
  %``The chiral and deconfinement aspects of the QCD transition,''
  Phys.\ Rev.\ D {\bf 85}, 054503 (2012).
%  [arXiv:1111.1710 [hep-lat]].
  %%CITATION = ARXIV:1111.1710;%%


%\cite{Bratovic:2012qs}
\bibitem{Bratovic:2012qs}
  N.~M.~Bratovic, T.~Hatsuda and W.~Weise,
 % ``Role of Vector Interaction and Axial Anomaly in the PNJL Modeling of the QCD Phase Diagram,''
  Phys.\ Lett.\ B {\bf 719}, 131 (2013).
  %[arXiv:1204.3788 [hep-ph]].
  %%CITATION = ARXIV:1204.3788;%%
  %19 citations counted in INSPIRE as of 05 Sep 2013

%\cite{Carignano:2010}
\bibitem{Carignano:2010}
  S.~Carignano, D.~Nickel, and M.~Buballa,
  %``Influence of vector interaction and Polyakov loop dynamics on inhomogeneous chiral symmetry breaking phases''
 Phys.\ Rev.\  D {\bf 82}, 054009 (2010).

%\cite{Kitazawa:2002bc}
\bibitem{Kitazawa:2002bc}
  M.~Kitazawa, T.~Koide, T.~Kunihiro and Y.~Nemoto,
  %``Chiral and color superconducting phase transitions with vector interaction in a simple model''
  Prog.\ Theor.\ Phys.\  {\bf 108}, 929 (2002).
%  [hep-ph/0207255, hep-ph/0307278].
  %%CITATION = HEP-PH/0207255,;%%

%\cite{Blaschke:2003cv}
\bibitem{Blaschke:2003cv}
  D.~Blaschke, M.~K.~Volkov and V.~L.~Yudichev,
  %``Coexistence of color superconductivity and chiral symmetry breaking within the NJL model,''
  Eur.\ Phys.\ J.\ A {\bf 17}, 103 (2003).
%  [hep-ph/0301065].
  %%CITATION = HEP-PH/0301065;%%

%\cite{Hatsuda:2006ps}
\bibitem{Hatsuda:2006ps}
  T.~Hatsuda, M.~Tachibana, N.~Yamamoto and G.~Baym,
  %``New critical point induced by the axial anomaly in dense QCD,''
  Phys.\ Rev.\ Lett.\  {\bf 97}, 122001 (2006).
%  [hep-ph/0605018].
  %%CITATION = HEP-PH/0605018;%%

%\cite{Blaschke:2004cc}
\bibitem{Blaschke:2004cc}
  D.~Blaschke, H.~Grigorian, A.~Khalatyan and D.~N.~Voskresensky,
  %``Exploring the QCD phase diagram with compact stars,''
  Nucl.\ Phys.\ Proc.\ Suppl.\  {\bf 141}, 137 (2005).
%  [hep-ph/0409116].
  %%CITATION = HEP-PH/0409116;%%

%\cite{Andronic:2009gj}
\bibitem{Andronic:2009gj}
  A.~Andronic, D.~Blaschke, P.~Braun-Munzinger, J.~Cleymans, K.~Fukushima, L.~D.~McLerran, H.~Oeschler and R.~D.~Pisarski {\it et al.},
  %``Hadron Production in Ultra-relativistic Nuclear Collisions: Quarkyonic Matter and a Triple Point in the Phase Diagram of  QCD,''
  Nucl.\ Phys.\ A {\bf 837}, 65 (2010).
%  [arXiv:0911.4806 [hep-ph]].
  %%CITATION = ARXIV:0911.4806;%%

\bibitem{Contrera:2007wu}
  G.~A.~Contrera, D.~Gomez Dumm and N.~N.~Scoccola,
  %``Nonlocal SU(3) chiral quark models at finite temperature: the role of the
  %Polyakov loop,''
  Phys.\ Lett.\  B {\bf 661} (2008) 113.
 % [arXiv:0711.0139 [hep-ph]].
  %%CITATION = PHLTA,B661,113;%%

\bibitem{Contrera:2010kz}
   G.~A.~Contrera, M.~Orsaria and N.~N.~Scoccola,
   %``Nonlocal Polyakov-Nambu-Jona-Lasinio model with wave function renormalization at finite temperature and chemical potential'',
   Phys.\ Rev.\  D {\bf 82}, 054026 (2010).
   %[arXiv:hep-ph/1006.4639]

%\cite{Horvatic:2010md}
\bibitem{Horvatic:2010md}
  D.~Horvatic, D.~Blaschke, D.~Klabucar, O.~Kaczmarek,
  %``Width of the QCD transition in a Polyakov-loop DSE model,''
  Phys.\ Rev.\ D {\bf 84}, 016005 (2011).
 % [arXiv:1012.2113 [hep-ph]].

%\cite{Radzhabov:2010dd}
\bibitem{Radzhabov:2010dd}
  A.~E.~Radzhabov, D.~Blaschke, M.~Buballa and M.~K.~Volkov,
  %``Nonlocal PNJL model beyond mean field and the QCD phase transition,''
  Phys.\ Rev.\ D {\bf 83}, 116004 (2011).
%  [arXiv:1012.0664 [hep-ph]].
  %%CITATION = ARXIV:1012.0664;%%

%\cite{Benic:2013eqa}
\bibitem{Benic:2013eqa}
  S.~Benic, D.~Blaschke, G.~A.~Contrera and D.~Horvatic,
  %``Medium induced Lorentz symmetry breaking effects in nonlocal PolyakovÐNambuÐJona-Lasinio models,''
  Phys.\ Rev.\ D {\bf 89}, no. 1, 016007 (2014).
%  doi:10.1103/PhysRevD.89.016007
%  [arXiv:1306.0588 [hep-ph]].
  %%CITATION = doi:10.1103/PhysRevD.89.016007;%%

%\cite{Rossner:2007ik}
\bibitem{Rossner:2007ik}
  S.~Roessner, T.~Hell, C.~Ratti and W.~Weise,
  %``The chiral and deconfinement crossover transitions: PNJL model beyond mean
  %field,''
Nucl.\ Phys.\  A {\bf 814}, 118 (2008).

\bibitem{Dumitru:2005ng}
A.~Dumitru, R.~D.~Pisarski and D.~Zschiesche,
Phys.\ Rev.\  D {\bf 72}, 065008 (2005).

\bibitem{Fukushima:2006uv}
K.~Fukushima and Y.~Hidaka,
Phys.\ Rev.\  D {\bf 75}, 036002 (2007).

\bibitem{Abuki:2008ht}
H.~Abuki, M.~Ciminale, R.~Gatto, G.~Nardulli and M.~Ruggieri,
Phys.\ Rev.\  D {\bf 77}, 074018 (2008);
H.~Abuki, R.~Anglani, R.~Gatto, G.~Nardulli and M.~Ruggieri,
Phys.\ Rev.\  D {\bf 78}, 034034 (2008).

%\citep{GomezDumm:2008sk}
\bibitem{GomezDumm:2008sk}
  D.~Gomez Dumm, D.~B.~Blaschke, A.~G.~Grunfeld and N.~N.~Scoccola,
  %``Color neutrality effects in the phase diagram of the PNJL model,''
  Phys.\ Rev.\  D {\bf 78}, 114021 (2008).
  %%CITATION = PHRVA,D78,114021;%%

\bibitem{Dexheimer:2009va}
  V.~A.~Dexheimer and S.~Schramm,
  %``A Novel Approach to Model Hybrid Stars,''
  Phys.\ Rev.\ C {\bf 81}, 045201 (2010).
  %%CITATION = ARXIV:0910.1312;%%

%\cite{Schaefer:2007pw}
\bibitem{Schaefer:2007pw}
  B.~-J.~Schaefer, J.~M.~Pawlowski and J.~Wambach,
  %``The Phase Structure of the Polyakov--Quark-Meson Model,''
  Phys.\ Rev.\ D {\bf 76}, 074023 (2007).
%  [arXiv:0704.3234 [hep-ph]].

\bibitem{Pagura:2012}
V.~Pagura, D.~G{\'o}mez Dumm and N.~N.~Scoccola,
%``Deconfinement and chiral restoration in non-local PNJL models at
%zero and imaginary chemical potential'',
Phys. Lett. B {\bf 707}, 76 (2012).
% [arXiv:hep-ph/1105.1739]

\bibitem{Noguera:2008prd78}
 S.~Noguera and N.~N.~Scoccola,
 %``Nonlocal chiral quark models with wavefunction renormalization:
 %sigma properties and pion-pion scattering parameters,''
 Phys.\ Rev.\  D {\bf 78}, 114002 (2008).
%[arXiv:hep-ph/0806.0818]

%\cite{Contrera:2012wj}
\bibitem{Contrera:2012wj}
  G.~A.~Contrera, A.~G.~Grunfeld and D.~B.~Blaschke,
  %``Phase diagrams in nonlocal Polyakov-Nambu-Jona-Lasinio models constrained by lattice QCD results,''
  Phys.\ Part.\ Nucl.\ Lett.\  {\bf 11}, 342 (2014).
 % [arXiv:1207.4890 [hep-ph]].
  %%CITATION = ARXIV:1207.4890;%%

%\cite{Blaschke:2007ri}
\bibitem{Blaschke:2007ri}
  D.~B.~Blaschke, D.~Gomez Dumm, A.~G.~Grunfeld, T.~Kl\"ahn and N.~N.~Scoccola,
  %``Hybrid stars within a covariant, nonlocal chiral quark model,''
  Phys.\ Rev.\  C {\bf 75}, 065804 (2007).
%  [arXiv:nucl-th/0703088].
  %%CITATION = PHRVA,C75,065804;%%

\bibitem{Fukushima2008}
K.~Fukushima,
 Phys.\ Rev.\ D {\bf 77}, 114028 (2008).

\bibitem{Dumm:2005}
D.~Gomez Dumm and N.~N.~Scoccola, Phys.\ Rev.\  C {\bf 72}, 014909 (2005).

%\cite{Friesen:2014mha}
\bibitem{Friesen:2014mha}
  A.~V.~Friesen, Y.~L.~Kalinovsky and V.~D.~Toneev,
  %``Vector interaction effect on thermodynamics and phase structure of QCD matter,''
  Int.\ J.\ Mod.\ Phys.\ A {\bf 30}, no. 16, 1550089 (2015).
%  doi:10.1142/S0217751X1550089X
%  [arXiv:1412.6872 [hep-ph]].
  %%CITATION = doi:10.1142/S0217751X1550089X;%%

\bibitem{Parappilly:2005ei}
  M.~B.~Parappilly, P.~O.~Bowman, U.~M.~Heller, D.~B.~Leinweber,
A.~G.~Williams and J.~B.~Zhang,
  %``Scaling behavior of quark propagator in full QCD,''
  Phys.\ Rev.\ D {\bf 73}, 054504 (2006).
  %[arXiv:hep-lat/0511007].
  %%CITATION = HEP-LAT 0511007;%%

%\cite{Allton:2003vx}
\bibitem{Allton:2003vx}
  C.~R.~Allton, S.~Ejiri, S.~J.~Hands, O.~Kaczmarek, F.~Karsch, E.~Laermann and C.~Schmidt,
  %``The Equation of state for two flavor QCD at nonzero chemical potential,''
  Phys.\ Rev.\ D {\bf 68}, 014507 (2003).
 % [hep-lat/0305007].
  %%CITATION = HEP-LAT/0305007;%%

%\cite{Fodor:2002km}
\bibitem{Fodor:2002km}
  Z.~Fodor, S.~D.~Katz and K.~K.~Szabo,
  %``The QCD equation of state at nonzero densities: Lattice result,''
  Phys.\ Lett.\ B {\bf 568}, 73 (2003).
 % [hep-lat/0208078].
  %%CITATION = HEP-LAT/0208078;%%

%\cite{AliKhan:2001ek}
\bibitem{AliKhan:2001ek}
  A.~Ali Khan {\it et al.} [CP-PACS Collaboration],
  %``Equation of state in finite temperature QCD with two flavors of improved Wilson quarks,''
  Phys.\ Rev.\ D {\bf 64}, 074510 (2001).
 % doi:10.1103/PhysRevD.64.074510
 % [hep-lat/0103028].
  %%CITATION = doi:10.1103/PhysRevD.64.074510;%%

\bibitem{Kaczmarek:2011zz}
  O.~Kaczmarek, F.~Karsch, E.~Laermann, C.~Miao, S.~Mukherjee, P.~Petreczky,
C.~Schmidt, W.~Soeldner and W. Unger,
  %``Phase boundary for the chiral transition in (2+1)-flavor QCD at small
 % values of the chemical potential,''
   Phys.\ Rev.\ D {\bf 83}, 014504 (2011).

%\cite{Bellwied:2015rza}
\bibitem{Bellwied:2015rza}
  R.~Bellwied, S.~Borsanyi, Z.~Fodor, J.~Günther, S.~D.~Katz, C.~Ratti and K.~K.~Szabo,
  %``The QCD phase diagram from analytic continuation,''
  arXiv:1507.07510 [hep-lat].
  %%CITATION = ARXIV:1507.07510;%%

%\cite{Bonati:2015bha}
\bibitem{Bonati:2015bha}
  C.~Bonati, M.~D'Elia, M.~Mariti, M.~Mesiti, F.~Negro and F.~Sanfilippo,
  %``Curvature of the chiral pseudocritical line in QCD: Continuum extrapolated results,''
  Phys.\ Rev.\ D {\bf 92}, no. 5, 054503 (2015)
 % [arXiv:1507.03571 [hep-lat]].
  %%CITATION = ARXIV:1507.03571;%%

%\cite{Andronic:2005yp}
\bibitem{Andronic:2005yp}
  A.~Andronic, P.~Braun-Munzinger and J.~Stachel,
  %``Hadron production in central nucleus-nucleus collisions at chemical freeze-out,''
  Nucl.\ Phys.\ A {\bf 772}, 167 (2006).
%  doi:10.1016/j.nuclphysa.2006.03.012
%  [nucl-th/0511071].
  %%CITATION = doi:10.1016/j.nuclphysa.2006.03.012;%%


%\cite{Ivanov:2012bh}
\bibitem{Ivanov:2012bh}
  Y.~B.~Ivanov,
  %``Baryon Stopping as a Probe of Deconfinement Onset in Relativistic Heavy-Ion Collisions,''
  Phys.\ Lett.\ B {\bf 721}, 123 (2013).
%  doi:10.1016/j.physletb.2013.02.038
%  [arXiv:1211.2579 [hep-ph]].
  %%CITATION = doi:10.1016/j.physletb.2013.02.038;%%

%\cite{Ivanov:2015vna}
\bibitem{Ivanov:2015vna}
  Y.~B.~Ivanov and D.~Blaschke,
  %``Robustness of the Baryon-Stopping Signal for the Onset of Deconfinement in Relativistic Heavy-Ion Collisions,''
  Phys.\ Rev.\ C {\bf 92}, no. 2, 024916 (2015).
%  doi:10.1103/PhysRevC.92.024916
%  [arXiv:1504.03992 [nucl-th]].
  %%CITATION = doi:10.1103/PhysRevC.92.024916;%%

%\cite{Sasaki:2006ws}
\bibitem{Sasaki:2006ws}
  C.~Sasaki, B.~Friman and K.~Redlich,
  %``Quark Number Fluctuations in a Chiral Model at Finite Baryon Chemical Potential,''
  Phys.\ Rev.\ D {\bf 75}, 054026 (2007)
%  doi:10.1103/PhysRevD.75.054026
%  [hep-ph/0611143].
  %%CITATION = doi:10.1103/PhysRevD.75.054026;%%

%\cite{Borsanyi:2013bia}
\bibitem{Borsanyi:2013bia}
  S.~Borsanyi, Z.~Fodor, C.~Hoelbling, S.~D.~Katz, S.~Krieg and K.~K.~Szabo,
  %``Full result for the QCD equation of state with 2+1 flavors,''
  Phys.\ Lett.\ B {\bf 730}, 99 (2014).
%  doi:10.1016/j.physletb.2014.01.007
%  [arXiv:1309.5258 [hep-lat]].
  %%CITATION = doi:10.1016/j.physletb.2014.01.007;%%

%\cite{Bazavov:2014pvz}
\bibitem{Bazavov:2014pvz}
  A.~Bazavov {\it et al.} [HotQCD Collaboration],
  %``Equation of state in ( 2+1 )-flavor QCD,''
  Phys.\ Rev.\ D {\bf 90}, 094503 (2014).
%  doi:10.1103/PhysRevD.90.094503
 % [arXiv:1407.6387 [hep-lat]].
  %%CITATION = doi:10.1103/PhysRevD.90.094503;%%

%\cite{Sasaki:2012bi}
\bibitem{Sasaki:2012bi}
  C.~Sasaki and K.~Redlich,
  %``An Effective gluon potential and hybrid approach to Yang-Mills thermodynamics,''
  Phys.\ Rev.\ D {\bf 86}, 014007 (2012).
%  doi:10.1103/PhysRevD.86.014007
%  [arXiv:1204.4330 [hep-ph]].
  %%CITATION = doi:10.1103/PhysRevD.86.014007;%%

%\cite{Ruggieri:2012ny}
\bibitem{Ruggieri:2012ny}
M.~Ruggieri, P.~Alba, P.~Castorina, S.~Plumari, C.~Ratti and V.~Greco,
%``Polyakov Loop and Gluon Quasiparticles in Yang-Mills Thermodynamics'',
Phys.\ Rev.\ D {\bf 86}, 054007 (2012).
%  arXiv:1204.5995 [hep-ph].
 % {\it Preprint} hep-ph/1204.5995.
  %%CITATION = ARXIV:1204.5995;%%

%\cite{Fukushima:2012qa}
\bibitem{Fukushima:2012qa}
K.~Fukushima and K.~Kashiwa,
% ``Polyakov loop and QCD thermodynamics from the gluon and ghost propagators'',
Phys.Lett. B {\bf 723}, 360 (2013).
%  arXiv:1206.0685 [hep-ph].
%  {\it Preprint} hep-ph/1206.0685.
  %%CITATION = ARXIV:1206.0685;%%

%\cite{Blaschke:2010ka}
\bibitem{Blaschke:2010ka}
  D.~B.~Blaschke, F.~Sandin, V.~V.~Skokov and S.~Typel,
  %``Accessibility of Color Superconducting Quark Matter Phases in Heavy-ion Collisions,''
  Acta Phys.\ Polon.\ Supp.\  {\bf 3}, 741 (2010).
%  [arXiv:1004.4375 [hep-ph]].
  %%CITATION = ARXIV:1004.4375;%%

%\cite{Blaschke:2013zaa}
\bibitem{Blaschke:2013zaa}
  D.~Blaschke, M.~Buballa, A.~Dubinin, G.~Roepke and D.~Zablocki,
  %``Generalized Beth--Uhlenbeck approach to mesons and diquarks in hot, dense quark matter,''
  Annals Phys.\  {\bf 348}, 228 (2014).
%  doi:10.1016/j.aop.2014.06.002
%  [arXiv:1305.3907 [hep-ph]].
  %%CITATION = doi:10.1016/j.aop.2014.06.002;%%

%\cite{Blaschke:2015nma}
\bibitem{Blaschke:2015nma}
  D.~Blaschke, A.~Dubinin and L.~Turko,
  %``Mott-hadron resonance gas and lattice QCD thermodynamics,''
  Phys.\ Part.\ Nucl.\  {\bf 46}, no. 5, 732 (2015).
%  doi:10.1134/S1063779615050093
%  [arXiv:1501.00485 [hep-ph]].
  %%CITATION = doi:10.1134/S1063779615050093;%%

%\cite{Dubinin:2015glr}
\bibitem{Dubinin:2015glr}
  A.~Dubinin, D.~Blaschke and A.~Radzhabov,
  %``Pion and kaon in the Beth-Uhlenbeck approach,''
  J.\ Phys.\ Conf.\ Ser.\  {\bf 668}, no. 1, 012052 (2016).
%  doi:10.1088/1742-6596/668/1/012052
%  [arXiv:1511.00338 [hep-ph]].
  %%CITATION = doi:10.1088/1742-6596/668/1/012052;%%

%\cite{Blaschke:2015sla}
\bibitem{Blaschke:2015sla}
  D.~Blaschke, A.~S.~Dubinin and D.~Zablocki,
  %``NJL model approach to diquarks and baryons in quark matter,''
  PoS BaldinISHEPPXXII {\bf }, 083 (2015);
  [arXiv:1502.03084 [nucl-th]].

\end{thebibliography}
\end{document}